\documentclass[12pt]{article}
\usepackage{a4,amsfonts,amsmath,bbm}
\begin{document}

% Definitions
\def\V#1{\mathcal{V}_{#1}}
\def\R#1{\mathcal{R}_{#1}}
\def\M#1{\mathcal{M}_{#1}}
\def\W{\mathcal{W}}
\def\Vbar#1{\overline{\mathcal{V}}_{#1}}
\def\Rbar#1{\overline{\mathcal{R}}_{#1}}
\def\Mbar#1{\overline{\mathcal{M}}_{#1}}

\def\vac{\mathit{\Omega}}
\def\logvac{\mathit{\omega}}
\def\one{\mathit{\phi}}
\def\logone{\mathit{\psi}}
\def\spin{a}
\def\index{\lambda}

\def\Nop{\hat{\mathcal{N}}}
\def\ndop{\hat{\delta}}    
\def\ndbarop{\hat{\bar{\delta}}}
\def\Lop{L}                
\def\Lbarop{\overline{L}}  
\def\Wop{W}
\def\Wbarop{\overline{W}{}}
\def\qop{\hat{q}}          
\def\hop{\hat{h}}         
\def\hbarop{\hat{\bar{h}}}
\def\Pop{\hat{\mathcal{P}}} 
\def\Pdaggerop{\Pop^{\dagger}} 
\def\Pbarop{\hat{\overline{\mathcal{P}}}}
\def\Pbardaggerop{\Pbarop{}^{\dagger}}
\def\del{\partial}
\def\Xop{{\cal O}}
\def\Xbarop{\overline{\Xop}}

\def\ket#1{\big |#1 \big \rangle}
\def\bra#1{\big \langle #1 \big |}
\def\braket#1#2{\big \langle #1 \big | #2\big \rangle}

\def\Rbs#1{\mathit{R}_{#1}}
\def\Vbs#1{\mathit{V}_{#1}}
\def\Xbs#1{\mathit{X}_{#1}}
\def\Ybs#1{\mathit{Y}_{#1}}
\def\Zbs#1{\mathit{Z}_{#1}}

\def\RNbs{\Rbs{0}}
\def\RObs{\Rbs{1}}
\def\VNbs{\Vbs{0}}
\def\VObs{\Vbs{1}}
\def\RNObs{\Rbs{01}}
\def\RONbs{\Rbs{10}}
\def\VOEbs{\Vbs{-1/8}}
\def\VTEbs{\Vbs{3/8}}
\def\XNbs{\Xbs{0}}
\def\XObs{\Xbs{1}}
\def\YNbs{\Ybs{0}}
\def\YObs{\Ybs{1}}
\def\bs{B}
\def\bsc{C}

\def\half{\frac{1}{2}}
\def\U{\mathcal{U}}
\def\Udagger{\U^{\dagger}}
\def\nn{{}}
\def\id{{\mathbbm 1}}
\def\ie{i.\,e.\ }
\def\wrt{w.\,r.\,t.\ }
\def\etal{{\em et al.}\ }
\def\ala{{\em {\`a} la}}
\def\eg{e.\,g.\ }
\def\character#1{\chi^{\ph{\V{}}}_{#1}}
\def\e{\mathrm{e}}
\def\imag{\mathrm{i}}
\def\ph#1{\phantom{#1}}
\def\phn{{\ph{0}}}
\def\phm{{\ph{-}}}

\def\bb#1{\mathord{\hbox{\boldmath$#1$}}}

\setlength{\textwidth}{14.4cm}
\setlength{\textheight}{21.9cm}
\setlength{\parindent}{0pt}

\begin{titlepage}

\begin{center}

\vspace*{-1.8cm}
{\hbox to \hsize{\hfill hep-th/0204154}}
{\hbox to \hsize{\hfill ITP-UH-08/02}}
\vspace*{1.8cm}

\vspace*{1.5cm}
{\Large \bf 
        Boundary States in ${\bb c=-2}$ Logarithmic\\[2mm] Conformal Field Theory}

\vspace*{1cm}
{       Andreas Bredthauer and
        Michael Flohr\footnote{\parbox[t]{12cm}{\tt 
        Andreas.Bredthauer@itp.uni-hannover.de\\
        Michael.Flohr@itp.uni-hannover.de}}}\\[1cm]

{\em	
        Institut f{\"u}r Theoretische Physik\\
        Universit{\"a}t Hannover\\
        Appelstra{\ss}e 2, 30167 Hannover, Germany}

\vspace*{1cm}
{\small April 18, 2002}

\vspace*{3cm}
{\small \bf Abstract}\\[0.5cm]

\parbox{12.5cm}
{\small
\hspace*{0.7cm}Starting from first principles, a constructive method is 
presented to obtain boundary states in conformal field theory. It is 
demonstrated that this method is well suited to compute the boundary states of
logarithmic conformal field theories. By studying the logarithmic conformal
field theory with central charge $c=-2$ in detail, we show that our method 
leads to consistent results. In particular, it allows to define  
boundary states corresponding to both, indecomposable representations as well 
as their irreducible subrepresentations.
}
\end{center}
\end{titlepage}

\newpage
\setcounter{page}{1}

%-----------------------------------------------------------
% section 1: Introduction
%-----------------------------------------------------------
\section{Introduction}

Conformal field theory in two dimensions \cite{BPZ:84} is undoubtedly one 
of the most important tools of modern theoretical physics with numerous
applications ranging from string theory to experimentally observed and
confirmed phenomena in condensed matter physics.
The last few years saw a renewed interest in conformal field theories, 
mainly arising from two different directions.
Much has been learned about conformal field theories on surfaces
with a boundary \cite{Ca:89,BePePeZu:00}. These are important since most 
real physical systems are finite in size and bounded. Surprisingly, 
boundary conformal field theory has also made its way into modern string 
theory where it presents a powerful tool for the computation of possible 
spectra of {\em D\/}-branes \cite{ReSch:98}. On the other hand, conformal 
field theories with logarithmic operators \cite{Gu:93} enjoy increasing 
attention, see \eg \cite{Ga:01,Fl:01} and references 
therein.
Again, phenomenology yields a good reason for them, since many condensed
matter systems, in particular systems involving disorder, possess density
fields with scaling dimension zero (this was first observed in the
treatment of dense polymers \cite{Sa:92}). Such fields are one cause for the
emergence of indecomposable representations and their associated
logarithmic operators. Other causes for logarithmically diverging
correlation functions which perhaps are more interesting for string
theory are, \eg twist fields in ghost systems (the problem was 
noted by {\nn V.~Knizhnik} \cite{Kn:87} as far back as 1987) or puncture 
operators in Liouville theory \cite{KoLe:98}.

Both directions in this field of research have reached a level of understanding
close to the one of rational conformal field theory. In ordinary conformal 
field theory, boundary states can be obtained via a formalism presented by 
{\nn N.~Ishibashi} \cite{Is:89}, which yields a constructive method to 
compute a basis of boundary states. In the case where the theory is rational 
{\nn J.\,L.~Cardy} \cite{Ca:89} gave a prescription how to obtain the physical 
set of boundary states relating the coefficients of the linear combinations to
the fusion coefficients of the bulk theory. Logarithmic conformal field 
theories share many properties with ordinary theories. Many 
notions of rationality can be generalized to such theories to a certain degree, 
such as characters, fusion rules, partition functions, see \eg 
\cite{Fl:96,GaKa:96,GaKa:961,Ro:96,Fl:97,GaKa:98,Ka:00,FjFuHwSeTi:02,Mi:01}.

A question which naturally arises is whether and how a boundary theory for
logarithmic conformal field theories can be obtained. {\nn I.\,I.~Kogan} and 
{\nn J.\,F.~Wheater} \cite{KoWh:00} were one of the first to discuss the 
question of boundary states and their effects in a $c=-2$ logarithmic 
conformal field theory. In a more recent work, {\nn S.~Kawai} and 
{\nn J.\,F.~Wheater} \cite{KaWh:01} studied boundary states of the same model 
by use of symplectic fermions. Further studies on $c=-2$ were conducted by 
{\nn Y.~Ishimoto} \cite{Ism:01}. Boundaries in the framework of logarithmic 
conformal field theories were also discussed by {\nn A.~Lewis} \cite{Le:00} as 
well as by {\nn S.~Moghimi-Araghi} and {\nn S.~Rouhani} \cite{MoRo:00}. 
The results of the former three works on boundaries in the $c=-2$ case are all
different, and partially contradictory. This demonstrates that the much
more complicated representation theory of logarithmic conformal field theories 
poses a major obstacle to a rigorous and consistent description of boundary 
states in these cases. For example, they naturally contain zero-norm states 
which cannot be neglected from the spectrum. This and their non-trivial inner 
structure make it impossible to get a normalized orthogonal basis of states
which usually is assumed in {\nn Ishibashi} like constructions of boundary
states.

In this paper, we modify the standard formalism \ala\/ {\nn Ishibashi}
in a suitable way to circumvent problems with non-normalizable states and
their interpretation. Our modified approach allows to construct boundary
states for logarithmic conformal field theories in a consistent way.
The paper will proceed as follows:

In section two, we present an algorithm to directly calculate a complete
basis of boundary states from first principles. It does not
make use of any assumptions on normalizability or orthogonality of a 
basis of states. Each finite level of a boundary state is computed
in a finite number of steps. As we will demonstrate, this algorithm is
naturally adepted to work on zero-norm states and to keep the inner
structure of indecomposable representations visible. Thus, our algorithm
should be appropriate for applications towards generic logarithmic conformal
field theories. Finally,
we show that our method is completely compatible to Ishibashi's approach,
yielding equivalent results for ordinary conformal field theories.

In the third section, we apply our method to the $c=-2$ logarithmic
conformal field theory, the best understood example of this species. We
concentrate on the $c=-2$ realization closest to the common notion of a
rational theory, where the maximally extended chiral symmetry algebra
is $\W(2,3,3,3)$. We obtain a complete set of 
boundary states in one-to-one correspondence to the representations
in this setting with respect to the extended 
symmetry algebra. 

Section four is devoted to a detailed discussion of various properties
of our boundary states. For instance, operators can be defined which
relate boundary states corresponding to (irreducible) subrepresentations of 
indecomposable Jordan block representations to the boundary states 
associated to the full indecomposable representations. 
We comment on a possible relation
between our set of boundary states and the structure of the unique local
logarithmic $c=-2$ conformal field theory constructed by {\nn M.\,R.~Gaberdiel}
and {\nn H.\,G.~Kausch} \cite{GaKa:98}. This is remarkable, since our set of 
boundary states shows a very similar structure with respect to the 
above-mentioned operators, although we did not start their construction from 
the local 
theory. We attempt to solve the problem of a degenerate metric of the natural
pairing in the space of boundary states by introducing additional so-called weak 
boundary states which serve as duals of some of the proper boundary states.

In section five, we concentrate on a subset of four boundary states that 
corresponds to the three-dimensional space spanned by the characters of the 
rational $c=-2$ theory. This set is well-defined and the induced metric is 
non-degenerate. We show that {\nn Cardy}'s formalism can be applied in this 
situation precisely as for ordinary rational conformal field theories, and that 
it yields consistent results, despite the fact that the ${\cal S}$ matrix does 
not diagonalize the fusion rules in the logarithmic conformal field theory case.

Section six repeats the above treatment for the case of the full
space of boundary states. Surprisingly, Cardy's formalism
still works to some extent. The partition functions are now related not
to the physical characters, but to functions forming a five-dimensional
representation of the modular group, which presumably can be interpreted
as torus amplitudes \cite{FlGa:02}. 
Interestingly, Cardy's formalism fails at precisely
the same point where a {\nn Verlinde} formula like computation of fusion 
coefficients within the five-dimensional representation of the modular group 
breaks down. A way out seems to be a limiting procedure, which eliminates the
weak boundary states introduced in section four. Unfortunately, the result of
this limit is a bit ambiguous and its physical interpretation
is not yet completely clear.

The paper concludes with a brief discussion, where we compare our
results with earlier works emphasizing the inner consistency of our
solution as well as the fact that it was derived without some of the
commonly made assumptions. These assumptions, although true for ordinary 
rational conformal field theory, typically do not hold in the
logarithmic case. Open questions and directions for future research are
also discussed.

%-----------------------------------------------------------
% section 2: The method
%-----------------------------------------------------------
\section{The method \label{section_method}}

In conformal field theories with boundaries, boundary conditions naturally
can be applied. These are implemented by boundary states.
Every such boundary state $\ket{\bs}$ to be compatible with conformal
invariance has to fulfill the condition that was analyzed 
and given by {\nn N.~Ishibashi}, namely
\begin{eqnarray}
  \left(\Lop_n-\Lbarop_{-n}\right)\ket{\bs}=0.
  \label{eqn:VirB}
\end{eqnarray}
This equation does not determine the boundary states completely because it can 
be read as an equation of motion for open string background in a closed string 
theory. Therefore, one usually analyzes boundary operators with respect to 
the maximally extended chiral symmetry algebra $\W$ with additional $N$ fields 
$\Wop^r,~r=1,\ldots,N$. The boundary state for this algebra has to obey in 
addition:
\begin{eqnarray}
  \left(\Wop_n^r-(-1)^{s_r}\Wbarop_{-n}^r\right)\ket{\bs}=0,
  \label{eqn:WB}
\end{eqnarray}
where $s_r$ labels the spin of the field $\Wop^r$. Let us denote a basis over
a given bulk representation module $\M{h}$ by 
\begin{eqnarray}
  \big\{\ket{l,n}\,\big|\,\,l=h,\,h+1,\ldots;\,n=1,\ldots,n_l\big\},
\end{eqnarray}   
where $l$ counts the levels beginning from the highest weight $h$ of the module 
and $n$ labels a suitable basis on each level. The metric $g_{mn}$ for this 
basis is given by the always well-defined and symmetric {\nn Shapovalov} forms 
(see \cite{Ro:96} for a prescription for logarithmic conformal field theories):
\begin{eqnarray}
  \delta_{ll^{\prime}}g_{mn}\equiv\braket{l,m}{l^{\prime},n}
  \equiv\lim_{z\rightarrow\infty}\lim_{w\rightarrow 0}z^{2l}
  \big\langle\phi_{l,m}(z)\phi_{l^\prime,n}(w)\big\rangle.
\end{eqnarray}
Here, $\phi_{l,m}$ is the field corresponding to the state $\ket{l,m}$. In 
ordinary conformal field theories, the basis states are chosen orthonormalized, 
\ie $g_{mn}\equiv\delta_{mn}$. However, to be more general we only demand a 
complete basis because the theory we especially consider contains zero-norm 
states. Thus, orthonormalization is not applicable. A boundary state can be 
written as a sum of product states 
$\ket{l,m;\bar{l},n}\equiv\ket{l,m}\otimes\overline{\ket{\bar{l},n}}$ of a 
holomorphic representation module $\M{h}$ and a formal anti-holomorphic module 
$\Mbar{\bar{h}}\,$:
\begin{eqnarray}
  \ket{\bs}=\sum_{l,m,\bar{l},n}c_{mn}^{\,l\,\bar{l}}
            \ket{l,m}\otimes\overline{\ket{\bar{l},n}}
           =\sum_{l,m,\bar{l},n}c_{mn}^{\,l\,\bar{l}}\ket{l,m;\bar{l},n}.
\end{eqnarray}
Note that we allow for $l\neq\bar{l}$ at this state. Of course, it will turn out 
that the solution contains only states of the type $\ket{l,m;l,n}$. Now consider 
equation \eqref{eqn:VirB}. The modes $\Lop_n$ obey the {\nn Virasoro} algebra:
\begin{eqnarray}
  \left[\Lop_m,\Lop_n\right]
  =\left(m-n\right)\Lop_{m+n}-\frac{c}{12}\left(m^3-m\right)\delta_{m+n,0}.
\end{eqnarray}  
{From} this one finds for $n\neq 2$:
\begin{eqnarray}
 \Lop_n=\frac{1}{n-2}\left[\Lop_{n-1},\Lop_1\right]\mbox{~~and~~}
 \Lop_{-n}=\frac{1}{2-n}\left[\Lop_{1-n},\Lop_{-1}\right].  
\end{eqnarray}
Thus, it is enough to check condition \eqref{eqn:VirB} for $n=-2,\ldots,2$ 
because the equation holds automatically for $|n|\geq 3$. For a boundary state 
that is built on two copies of the same representation module in its holomorphic 
and anti-holomorphic part this statement is equivalent to the requirement that 
\eqref{eqn:VirB} holds for $n=0,1,2$ while the coefficients are chosen 
symmetrically in $m$ and $n$, \ie 
$c_{mn}^{\,l\,\bar{l}}=c_{nm}^{\,l\,\bar{l}}$.

Imagine that $\Lop_0$ and $\Lbarop_0$ are given in {\nn Jordan} form and let us 
decompose them into their diagonal part $\hop$ and their off-diagonal part 
$\ndop$, such that $\Lop_0=\hop+\ndop$ and $\Lbarop_0=\hbarop+\ndbarop$. 
Equation \eqref{eqn:VirB} then reads:
\begin{align}
  \begin{split}
    0&=\left(\Lop_0-\Lbarop_0\right)\ket{\bs} \\[2ex]
     &=\left(\hop-\hbarop+\ndop-\ndbarop\right)
       \sum_{l,m,\bar{l},n}c_{mn}^{\,l\,\bar{l}}\ket{l,m;\bar{l},n} \\
     &=\sum_{l,m,\bar{l},n}c_{mn}^{\,l\,\bar{l}}
       \left(l-\bar{l}+\ndop-\ndbarop\right)\ket{l,m;\bar{l},n}.
  \end{split} 
  \label{eqn:L0}
\end{align}
Since the basis states $\ket{l,m;\bar{l},n}$ are linearly independent, it hereby 
follows that the off-diagonal part $(\ndop-\ndbarop)\ket{\bs}$ has to vanish 
and, on the other hand, $l=\bar{l}$:
\begin{eqnarray}
  \ket{\bs}=\sum_{l,m,n}c_{mn}^{\,l}\ket{l,m;l,n}.
  \label{eqn:bs-sum}
\end{eqnarray}
This is not quite unexpected. It is part of the result given by {\nn N.~Ishibashi} 
for ordinary conformal field theories. In the same manner the treatment of the 
$n=1$ case results in the following equations:
\begin{align}
 \begin{split}
  0&=\left(\Lop_1-\Lbarop_{-1}\right)\ket{\bs} \\[1.5ex]
   &=\left(\Lop_1-\Lbarop_{-1}\right)\sum_{l,m,n}c_{mn}^{\,l}\ket{l,m;l,n} \\
   &=\sum_{l,m,n}
     c_{mn}^{\,l}\Big(\sum_a\alpha^{\,l\,m}_{\ph{\,l\,m}a}\ket{l-1,a;l,n}
     -\sum_b\beta^{\,l\,n}_{\ph{\,l\,n}b}\ket{l,m;l+1,b}\Big)  \\
   &=\sum_{l,m,n}\left(\alpha^{\,l\,a}_{\ph{\,l\,a}m}c_{an}^{\,l}-
     c_{mb}^{\,l-1}\beta^{\,l-1\,b}_{\ph{\,l-1\,b}n}\right)\ket{l-1,m;l,n}.
 \end{split}
 \label{eqn:L1}
\end{align}
Here, $\alpha$ and $\beta$ denote the coefficients in the expansion of 
$\Lop_1\ket{l,m}$ and $\overline{\Lop_{-1}\ket{l,n}}$ with respect to the 
basis $\ket{l-1,a}$ and $\overline{\ket{l+1,b}}$:
\begin{eqnarray}
  \Lop_1\ket{l,m}=\alpha^{\,l\,m}_{\ph{\,l\,m}a}\ket{l-1,a}\mbox{~~and~~}
  \overline{\Lop_{-1}\ket{l,n}}=\beta^{\,l\,n}_{\ph{\,l\,n}b}
  \overline{\ket{l+1,b}}.
\end{eqnarray}
We used {\nn Einstein}'s summing convention for the indices $a$ and $b$ in the 
last line. Because zero can only be created trivially out of the basis states 
all coefficients have to vanish identically, level by level:
\begin{eqnarray}
  \alpha^{\,l\,a}_{\ph{\,l\,a}m}c_{an}^{\,l}
  -c_{mb}^{\,l-1}\beta^{\,l-1\,b}_{\ph{\,l-1\,b}n}=0.
\end{eqnarray}
For $n=2$ it follows analogously that
\begin{eqnarray}
  \varrho^{\,l\,a}_{\ph{\,l\,a}m}c_{an}^{\,l}
  -c_{mb}^{\,l-2}\sigma^{\,l-2\,b}_{\ph{\,l-2\,b}n}=0.
 \label{eqn:L2}
\end{eqnarray}
Again, we introduced the expansion coefficients $\varrho$ and $\sigma$ for the 
states $\Lop_2\ket{l,m}$ and $\overline{\Lop_{-2}\ket{l,n}}$. The condition 
\eqref{eqn:WB} for the extended symmetry algebra $\W$ has to be treated in the 
same way. These equations reduce to a finite set as well. The additional fields 
in the extended chiral symmetry algebra are primary with respect to the 
energy-momentum tensor. Let $h$ be the conformal weight of the field $\Wop^r$:
\begin{eqnarray}
  \big[\Lop_n^{\ph{r}},\Wop_m^r\big]=\left((h-1)m-n\right)\Wop_{m+n}^r.
\end{eqnarray}
From this we learn:
\begin{eqnarray}
  \Wop_n^{\spin}=\left\{\begin{array}{ll}
    \frac{1}{(h-1)n}\big[\Lop_n^{\ph{r}},\Wop_0^{r}\big]&(n\neq 0,~h\neq 1) 
    \vspace{1.5ex} \\
    \big[\Lop_{n-1}^{\ph{r}},\Wop_1^{r}\big]&(h=1)
  \end{array}\right.\,.  
\end{eqnarray}
Therefore, it is enough to check \eqref{eqn:WB} for $n=0$ (or $n=0,\,\pm 1$ if 
$h=1$), since the remaining conditions are treated implicitly with the help of 
\eqref{eqn:VirB}:
\begin{eqnarray}
  \left(\Wop_0^{r}-(-1)^{s_r}\Wbarop_0^{r}\right)\ket{\bs}=0,~r=1,\ldots,N.
  \label{eqn:generalWB}
\end{eqnarray}
This leads to additional $N+2k$ equations, where $N$ is the total number of 
fields additional to the {\nn Virasoro} field and $k$ is the number of fields of 
conformal weight 1 among these. In particular, in the special case of the 
$\W(2,3,3,3)$ algebra in the $c=-2$ theory the three additional fields 
$\Wop^{\spin}$ are spin-3 fields and equation \eqref{eqn:generalWB} reduces to 
the three equations 
\begin{eqnarray}
  \left(\Wop_0^{\spin}+\Wbarop_0^{\spin}\right)\ket{\bs}=0,
  \label{eqn:extAlg}
\end{eqnarray}
where $\spin$ is the spinor index of $su(2)$ and takes three different values. 
By solving all the conditions for the coefficients $c_{mn}^{\,l}$ we are able to 
find a complete basis of boundary states. We state that after having derived the 
coefficients for the first three levels without any inconsistencies, there are 
no contradictions occurring on higher levels. One can be sure that the state 
really exists in the way that it is a well-defined solution of \eqref{eqn:VirB} 
and \eqref{eqn:WB}. Indeed, this is the case for the $c=-2$ rational logarithmic 
conformal field theory considered here and we will show the existence of the 
states we derive explicitly later on. In particular, this allows to calculate
the coefficients $c^{\,l}_{mn}$ for any arbitrary given finite level $l$ with a
finite number of steps. Given an arbitrary set of representation 
modules $\M{h}^1,~m=1,\ldots,n$ that build a {\nn Jordan} block of rank $n$ in 
the $\Lop_0$ mode, the number of boundary states built on 
$\M{h}^1\otimes\overline{\M{}}{}_{h}^1$ derived by our method is given by $n$. 
One state is built on the whole representation module and one each for the 
subrepresentations, based only on the product states of these 
subrepresentations, respectively. This is seen very easily, namely let $\M{h}^1$
be a subrepresentation in $\M{h}^1$ that behaves as a true 
highest weight representation if taken for itself. Acting with annihilators on 
the states in $\M{h}^m$ will never lead to dependencies in the equations 
\eqref{eqn:VirB} and \eqref{eqn:WB}.

It is obvious that in ordinary conformal field theories this formalism 
reproduces the usual {\nn Ishibashi} results: Let $\{\ket{l,m}\}$ be an 
orthonormal basis of an irreducible module $\V{}$. The corresponding 
{\nn Ishibashi} boundary state reads:
\begin{eqnarray}
  \ket{\Vbs{}}=\sum_{l,m}\ket{l,m}\otimes\overline{\U\ket{l,m}}.
  \label{eqn:Isbs}
\end{eqnarray}
$\U$ is an anti-unitary operation with the property that it acts on the modes of 
the extended chiral algebra as $\Wop^r_n\U=(-1)^{s_r}\U\Wop^r_n$ and commutes 
with the {\nn Virasoro} modes. Since the state given in \eqref{eqn:Isbs} 
satisfies the two equations \eqref{eqn:VirB} and \eqref{eqn:WB}, it has to 
fulfill the coefficient equations \eqref{eqn:L0}, \eqref{eqn:L1}, and 
\eqref{eqn:L2} as well as the corresponding ones for the modes of the extended 
algebra \eqref{eqn:extAlg}. This means that we can construct $\ket{\Vbs{}}$ by 
applying our formalism. The difference is that we do not make explicit use of 
the anti-unitary operation $\U$. By splitting off $\U$ by hand afterwards, we 
reproduce exactly the well-known results of ordinary conformal field theories. 
On the other hand we introduced a generalized procedure that allows us to take 
care of \eg zero-norm states in logarithmic conformal field theories. This 
generalization does not make any use of the properties of the states themselves 
other than the expansion with respect to an arbitrary but fixed basis.

%-----------------------------------------------------------
% section 3: Boundary states in rational c=-2 logarithmic
%            conformal field theory
%-----------------------------------------------------------
\section{Boundary states in ${\bb c=-2}$ rational logarithmic conformal 
field theory}

Here, we apply the method introduced in the previous section to the rational 
$c=-2$ logarithmic conformal field theory. {\nn M.\,R.~Gaberdiel} and 
{\nn H.\,G.~Kausch} spent a lot of work in the analysis of this (bulk) theory 
in \cite{GaKa:96,GaKa:961,GaKa:98,Ka:95}. {\nn F.~Rohsiepe} examined the 
physical characters of the representation modules that form the three-dimensional 
representation of the modular group \cite{Ro:96}. The torus amplitudes on the 
other hand seem to be related to a five-dimensional representation 
\cite{FlGa:02} that was analyzed in \cite{Fl:97}. The latter representation 
contains the smaller one as a subrepresentation. The theory contains a 
$\W(2,3,3,3)$ triplet algebra which is generated by the {\nn Virasoro} modes 
$\Lop_n$ and the modes $\Wop_n^{\spin}$ of a triplet of spin-3 fields. With the 
help of two quasi-primary normal-ordered fields 
$\Lambda=\;:\!\Lop^2\!\!:\!-3/10\,\del^2\Lop$ and 
$V^{\spin}=\;:\!\Lop\Wop^{\spin}\!\!:\!-3/14\,\del^2\Wop^{\spin}$ the 
commutation relations read:
\begin{align}
  \big[\Lop_m,\Lop_n\big]\,=&(m-n)\Lop_{m+n}-\frac{1}{6}
    \left(m^3-m\right)\delta_{m+n,0}, \nonumber \\
  \big[\Lop_m^{\ph{r}},\Wop_n^a\big]=&(2m-n)\Wop_{m+n}^a, \nonumber \\
  \big[\Wop_m^a,\Wop_n^b\big]=
    &g^{ab}\left(2\left(m-n\right)\Lambda_{m+n}+\frac{1}{20}\left(m-n\right)
     \left(2m^2+2n^2-mn-8\right)\Lop_{m+n}\right. \nonumber \\
    &\ph{g^{ab}\Big(}\left.-\frac{1}{120}m\left(m^2-1\right)
     \left(m^2-4\right)\delta_{m+n,0}\right) \\
    &\vspace*{1cm}+f^{ab}_c\left(\frac{5}{14}\left(2m^2+2n^2-3mn-4\right)
     \Wop^c_{m+n}+\frac{12}{5}V^c_{m+n}\right). \nonumber
  \label{eqn:commutation-relations}
\end{align}
Here, $g^{ab}$ is the metric and $f^{ab}_c$ are the structure constants of 
$su(2)$. For our further discussion it is suitable to choose a 
{\nn Cartan}-{\nn Weyl} basis for $su(2)$ which reads $\Wop^0$, $\Wop^{\pm}$, 
such that the metric is given by $g^{00}=1$, $g^{+-}=g^{-+}=2$ and the 
non-vanishing structure constants read $f^{0\pm}_{\pm}=-f^{\pm 0}_{\pm}=\pm 1$ 
and $f^{+-}_0=-f^{-+}_0=2$. Commutators involving the operators $\ndop$ and 
$\hop$ read
\begin{eqnarray}
  \big[\hop,\Xop_n\big]=\big[\Lop_0,\Xop_n\big]&\mbox{and}&
  \big[\ndop,\Xop_n\big]=0,
  \label{eqn:delta-h-commutators}
\end{eqnarray}
where $\Xop_n$ is any mode of the algebra. The theory contains four irreducible 
representations, two singlet representations, namely the vacuum representation 
$\V{0}$ and $\V{-1/8}$ with highest weight states $\vac$ at $h=0$ and $\mu$ at 
$h=-1/8$, respectively, and two doublet representations $\V{1}$ and $\V{3/8}$ 
with highest weight states at $h=1$ and $h=3/8$. Furthermore there exist two 
reducible but indecomposable representations: $\R{0}$ is generated by a cyclic 
state $\logvac$ at level 0 that builds a {\nn Jordan} block in $\Lop_0$ together 
with the vacuum highest weight state $\vac$ of $\V{0}$ and $\R{1}$ is generated 
by a doublet of level 1 cyclic states $\logone^{\pm}$ that form {\nn Jordan} 
blocks together with the highest weight states $\one^{\pm}$ of the 
representation $\V{1}$. These cyclic states have the property that they 
themselves are no highest weight states, \ie 
$\xi^{\pm}\equiv -\half\Lop_1\logone^{\pm}$ is not zero. The 
representations $\V{0}$ and $\V{1}$ are subrepresentations of the modules 
$\R{0}$ and $\R{1}$, respectively. Due to the fact that the highest occurring 
weight in both of these indecomposable representations is $h=0$, \ie their 
spectra are bounded from below as in irreducible representations, these 
representations are also called generalized highest weight representations. It 
follows that the states in the two (sub-)representations $\V{0}$ and $\V{1}$ are 
zero-norm states \cite{Ro:96}. Furthermore, $\R{0}$ contains two 
subrepresentations of type $\V{1}$ built on the two doublet states $\Psi_1^\pm$ 
and $\Psi_2^\pm$ (in this we follow the conventions of \cite{GaKa:961}):
\begin{eqnarray} 
 \begin{array}{rcl@{\qquad}rcl}
  \Psi_1^+&\!\!\!=\!\!\!&\Wop_{-1}^+\logvac,&
  \Psi_2^+&\!\!\!=\!\!\!&\left(\Wop_{-1}^0+\half\Lop_{-1}^{\phn}\right)\logvac, 
  \vspace{1ex} \\
  \Psi_1^-&\!\!\!=\!\!\!&\left(-\Wop_{-1}^0+\half\Lop_{-1}^{\phn}\right)\logvac,
 &\Psi_2^-&\!\!\!=\!\!\!&\Wop_{-1}^-\logvac.
 \end{array} 
\end{eqnarray}
The structure of the indecomposable modules $\R{0}$ and $\R{1}$ can be drawn 
schematically:
\begin{displaymath}
  \begin{array}{c@{\qquad\qquad}c}
    \begin{picture}(150,120)(-10,-20)
      \put(0,0){\vbox to 0pt
        {\vss\hbox to 0pt{\hss$\bullet$\hss}\vss}}
      \put(129,0){\vbox to 0pt
        {\vss\hbox to 0pt{\hss$\bullet$\hss}\vss}}
      \put(40,60){\vbox to 0pt
        {\vss\hbox to 0pt{\hss$\bullet$\hss}\vss}}
      \put(89,60){\vbox to 0pt
        {\vss\hbox to 0pt{\hss$\bullet$\hss}\vss}}
      \put(37,56){\vector(-2,-3){34}}
      \put(85,57){\vector(-3,-2){81}}
      \put(125,3){\vector(-3,2){81}}
      \put(126,5){\vector(-2,3){34}}
      \put(124,0){\vector(-1,0){119}}
      \put(-5,-15){$\vac$}
      \put(124,-15){$\logvac$}
      \put(35,70){$\Psi_1^\pm$}
      \put(84,70){$\Psi_2^\pm$}
    \end{picture}
    &
    \begin{picture}(140,120)(-10,-20)
      \put(0,60){\vbox to 0pt
        {\vss\hbox to 0pt{\hss$\bullet$\hss}\vss}}
      \put(124,60){\vbox to 0pt
        {\vss\hbox to 0pt{\hss$\bullet$\hss}\vss}}
      \put(62,0){\vbox to 0pt
        {\vss\hbox to 0pt{\hss$\bullet$\hss}\vss}}
      \put(119,60){\vector(-1,0){114}}
      \put(57,3){\vector(-1,1){54}}
      \put(121,57){\vector(-1,-1){54}}
      \put(-5,70){$\one^\pm$}
      \put(119,70){$\logone^\pm$}
      \put(55,-15){$\xi^\pm$}
    \end{picture}
    \\
    {\cal R}_0 & {\cal R}_1
  \end{array}
\end{displaymath}
\refstepcounter{figure} \label{fig:R0-and-R1}
\begin{center}{\em figure \ref{fig:R0-and-R1}: 
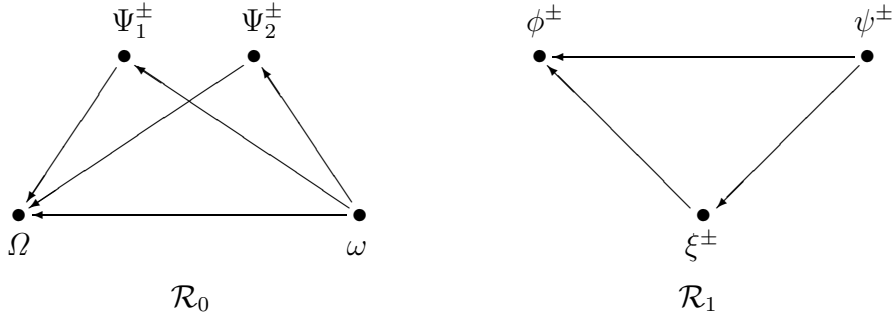
  Generalised highest weight modules $\R{0}$ and $\R{1}$}
\end{center}
The points in figure \ref{fig:R0-and-R1} refer to the states on which the 
different (sub-)re\-pre\-sen\-ta\-tions are built and the lines denote the 
action of the $\W$ algebra. {\nn Y.~Ishimoto} showed that we can choose the 
metric in these two representations to be given by \cite{Ism:01}:
\begin{eqnarray}
  \begin{array}{ccl@{\qquad}ccl@{\qquad}ccl}
    \braket{\vac}{\vac}&\!\!\!=\!\!\!&0,&
    \braket{\vac}{\logvac}&\!\!\!=\!\!\!&1,&
    \braket{\logvac}{\logvac}&\!\!\!=\!\!\!&d, \vspace{2ex} \\
    \braket{\one^+}{\one^-}&\!\!\!=\!\!\!&0,&
    \braket{\one^+}{\logone^-}&\!\!\!=\!\!\!&-1,&
    \braket{\logone^+}{\logone^-}&\!\!\!=\!\!\!&-t.
  \end{array}
  \label{eqn:metricR}
\end{eqnarray}
Here, $d$ and $t$ are in principal arbitrary real numbers. This fixes the metric 
completely. Applying the previously introduced method we find ten boundary 
states:
\begin{itemize}
\item A single state $\ket{\VOEbs}$ for the pairing 
      $\V{-1/8}\otimes\overline{\V{-1/8}}$
\item Another state $\ket{\VTEbs}$ for $\V{3/8}\otimes\overline{\V{3/8}}$
\end{itemize}
These states are the usual {\nn Ishibashi} states for $\V{-1/8}$ and $\V{3/8}$. 
The situation is different for states built on $\R{0}\otimes\overline{\R{0}}$. 
We find two independent coefficients 
$c_{\vac\vac}^{\phn}\equiv c_{\vac\vac}^{\,0}$ and 
$c_{\vac\logvac}^{\phn}\equiv c_{\vac\logvac}^{\,0}$ and hence two different 
solutions. This is exactly what we stated above. There exists one state for the 
complete module $\R{0}$ and one for the submodule $\V{0}$\,:
\begin{eqnarray}
  \begin{array}{rcl}
    \ket{\RNbs^{}}&\!\!\!=\!\!\!&\ket{c_{\vac\vac}=-d,~c_{\vac\logvac}=1}, 
    \vspace{1ex} \\
    \ket{\VNbs^{}}&\!\!\!=\!\!\!&\ket{c_{\vac\vac}=1,~c_{\vac\logvac}=0}.
  \end{array}
\end{eqnarray}
On the right hand side, the boundary states are defined via the given choice 
of the two free parameters, fixing all other coefficients in \eqref{eqn:bs-sum}.
In 
$\ket{\RNbs}$ we choose the parameter $c_{\vac\vac}=-d$ to be convenient for our 
further discussion. Remember that $d$ is a structure constant fixed in the 
metric \eqref{eqn:metricR}. Analogously, for $\R{1}\otimes\overline{\R{1}}$ one 
derives:
\begin{eqnarray}
  \begin{array}{rcl}
    \ket{\RObs}&\!\!\!=\!\!\!&\ket{c_{\xi^+\xi^-}=-t,~c_{\one^+\logone^-}=1},
    \vspace{1ex} \\
    \ket{\VObs}&\!\!\!=\!\!\!&\ket{c_{\xi^+\xi^-}=1,~c_{\one^+\logone^-}=0},
  \end{array}
\end{eqnarray}
where again $c_{\xi^+\xi^-}=-t$ is chosen for convenience. Note that here, the 
coefficients are antisymmetric with respect to interchanging the 
$su(2)$-spin indices. 

Let us expand the states $\ket{\RNbs}$ and $\ket{\RObs}$ into the sums 
\eqref{eqn:bs-sum} and re-introduce the usual anti-unitary operation $\U$ which 
acts as $\Wop^\pm_n\U=-\U\Wop_n^\mp$ and $\Wop^0_n\U=-\U\Wop^0_n$. Finally, we 
introduce a coefficient matrix $\gamma$ that is implicitly defined by 
$\gamma\cdot(1\otimes\overline{\U})\equiv c$. The two states take the 
following explicitly written out form:
\begin{eqnarray}
  \ket{\Rbs{\index}}=\sum_{l,m,n}\gamma^{\index\,\,l}_{\ph{\index}\,mn}
  \ket{l,m}\otimes\overline{\U\ket{l,n}},~~\index=0,1.
\end{eqnarray}
It turns out that for $\index=0,\,1$ the coefficient matrices $\gamma^\index$ 
are given by the inverse metrics on the corresponding representations 
$\R{\index}$. In this sense, the states $\ket{\Rbs{\index}}$ can be called 
{\em generalized\/} {\nn Ishibashi} states. They are defined free of 
contradictions, since:
\begin{eqnarray}
  0&\!\!\!=\!\!\!&\left(\Xop_n\pm\Xbarop_{-n}\right)\ket{\Rbs{\index}} 
      \nonumber \\
   &\!\!\!=\!\!\!&\bra{l_1,a}\otimes\overline{\U\bra{l_1,b}}
      \left(\Xop_n\pm\Xbarop_{-n}\right)
      \sum_{l,m,n}\gamma^{\index\,\,l}_{\ph{\index}\,mn}
      \ket{l,m}\otimes\overline{\U\ket{l,n}} \\
   &\!\!\!=\!\!\!&\sum_{l,m,n}\left(
      \bra{l_1,a}\Xop_n\gamma^{\index\,\,l}_{\ph{\index}\,mn}\ket{l,m}
      \braket{l,m}{l_2,b}-  
      \bra{l_1,a}\gamma^{\index\,\,l}_{\ph{\index}\,mn}\ket{l,m}
      \bra{l,m}\Xop_n\ket{l_2,b}\right)\nonumber\\
   &\!\!\!=\!\!\!&\bra{l_1,a}\left[\Xop_n,\id^{\index}\right]\ket{l_2,b} 
   \nonumber
\end{eqnarray}
for the modes $\Xop_n$ of the chiral algebra. The operator $\id^{\index}$ 
defined by
\begin{eqnarray}
  \id^{\index}\equiv\sum_{l,m,n}
  \gamma^{\index\,\,l}_{\ph{\index}\,mn}\ket{l,m}\!\bra{l,n}
\end{eqnarray}
is the projector onto the representation module $\R{\index}$. Indeed,
\begin{eqnarray}
  \left(\id^{\index}\right)^2&\!\!\!=\!\!\!&\sum_{l,m,n}\sum_{k,a,b}
    \ket{l,m}\gamma^{\index\,\,l}_{\ph{\index}\,mn}
    \underbrace{\braket{l,n}{k,a}}_{\delta_{lk}g_{na}}
    \gamma^{\index\,k}_{\ph{\index}\,ab}\bra{k,b} \nonumber \\
  &\!\!\!=\!\!\!&\sum_{l,m,n}\sum_{k,a,b}
    \ket{l,m}\delta^{\ph{\index}}_{lk}\delta^{\ph{\index}}_{ma}
    \gamma^{\index\,k}_{\ph{\index}\,ab}\bra{k,b} \\
  &\!\!\!=\!\!\!&\sum_{l,m,b}\gamma^{\index\,\,l}_{\ph{\index}\,mb}
    \ket{l,m}\!\bra{l,b}\,=\,\id^{\index}.
  \nonumber
\end{eqnarray}
Hence, it commutes with the action of the algebra.

In ordinary conformal field theories there are no boundary states based on 
product states of different representations in their holomorphic and 
anti-ho\-lo\-mor\-phic part because the weights of two different representations 
are usually disjunct sets. We have the representations $\R{0}$ and $\R{1}$ which 
contain the same weights and even their characters are equal \cite{GaKa:961}. 
Indeed, we find another two doublets of boundary states for the combinations 
$\R{0}\otimes\overline{\R{1}}$ and $\R{1}\otimes\overline{\R{0}}$\,:
\begin{eqnarray}
  \begin{array}{rcl}
    \ket{\RNObs^\pm}&\!\!\!=\!\!\!&\ket{c_{\vac\xi^\pm}=1}, \vspace{1ex} \\
    \ket{\RONbs^\pm}&\!\!\!=\!\!\!&\ket{c_{\xi^\pm\vac}=1}.
  \end{array}
\end{eqnarray}
To summarize the results, we could identify six solutions for the boundary 
states that have a one-to-one correspondence to the representation modules and 
two doublet solutions that relate the two generalized highest weight 
representations to each other. These span the space of all possible boundary states 
in this theory.

%-----------------------------------------------------------
% section 4: Properties of the solution
%-----------------------------------------------------------
\section{Properties of the solution \label{section_discussion}}

This section deals with the analysis of the properties of the states derived in 
the last paragraph. A common problem in the treatment of boundary states is that 
scalar products naturally diverge unless they are equal to zero because these 
states are infinite sums over tensor products of bulk states possessing finite 
scalar products. Let us introduce the operator
\begin{eqnarray}
  \Nop\equiv\ndop+\ndbarop.
\end{eqnarray}
Recall that we have $\Lop_0=\hop+\ndop$ and that in the $c=-2$ theory $\Lop_0$ 
has {\nn Jordan} blocks of dimension 2 at maximum for the $\R{0}$ and $\R{1}$ 
representations. Thus, $\ndop^2=0$, and it follows that the operator $\Nop$ is 
nilpotent of degree three: $\Nop^3=0$. Given this operator it is clear by use of 
\eqref{eqn:delta-h-commutators} that if $\ket{\bs}$ is a boundary state then 
$\Nop\ket{\bs}$ either vanishes or is a boundary state itself. We find that
\begin{eqnarray}
  \ket{\Vbs{\index}}=\half\Nop\ket{\Rbs{\index}}=-\del\ket{\Rbs{\index}},
  \mbox{~~$\index$=0,\,1},
\end{eqnarray}
where $\del\equiv\del_d$ if acting on $\ket{\RNbs}$ and $\del\equiv -\del_t$ if 
acting on $\ket{\RObs}$. This shows that the two states $\ket{\Vbs{\index}}$ are 
well-defined. The structure is very similar to the bulk theory:
\vspace*{-0.5cm}
\begin{displaymath}
  \begin{array}{c@{\qquad\qquad}c}
    \begin{picture}(150,60)(-10,-20)
      \put(0,0){\vbox to 0pt
        {\vss\hbox to 0pt{\hss$\bullet$\hss}\vss}}
      \put(129,0){\vbox to 0pt
        {\vss\hbox to 0pt{\hss$\bullet$\hss}\vss}}
      \put(124,0){\vector(-1,0){119}}
      \put(-10,-15){$\vac,~\one$}
      \put(119,-15){$\logvac,~\logone$}
      \put(60,7){$\ndop$}
    \end{picture}
  &
    \begin{picture}(150,60)(-10,-20)
      \put(0,0){\vbox to 0pt
        {\vss\hbox to 0pt{\hss$\bullet$\hss}\vss}}
      \put(129,0){\vbox to 0pt
        {\vss\hbox to 0pt{\hss$\bullet$\hss}\vss}}
      \put(124,0){\vector(-1,0){119}}
      \put(-5,-15){$\ket{\Vbs{\index}}$}
      \put(124,-15){$\ket{\Rbs{\index}}$}
      \put(60,7){$\frac{1}{2}\Nop$}
    \end{picture}
    \\
    \mbox{\em bulk states} & \mbox{\em boundary states}
  \end{array}
\end{displaymath}
\refstepcounter{figure} \label{fig:N_eq_delta}
\vspace*{-0.5cm}
\begin{center}
{\em figure \ref{fig:N_eq_delta}: Equivalence of $\ndop$ and $\Nop$}
\end{center}
Every boundary state $\ket{\bs}$ satisfies
\begin{eqnarray}
  \Nop^2\ket{\bs}=\del^2\ket{\bs}=0. \label{eqn:Nsquare_zero}
\end{eqnarray}
This can be seen if one remembers equation \eqref{eqn:L0} where we showed that a
boundary state $\ket{\bs}$ satisfies $(\ndop-\ndbarop)\ket{\bs}=0$. Using the 
nilpotency of $\ndop$ and $\ndbarop$ one gains
\begin{eqnarray}
  0=(\ndop-\ndbarop)^2\ket{\bs}=-2\ndop\ndbarop\ket{\bs}=-\Nop^2\ket{\bs}.
\end{eqnarray}  
Every partition function in a theory with boundaries can be written as a linear 
sum of the {\nn Virasoro} characters 
$\character{i}\!(q)=q^{-c/24}\,\mathrm{tr}_i\,q^{\Lop_0}$ of its representations 
$i$. For example, the partition function of a theory with two boundaries and 
boundary conditions $\alpha$ and $\beta$, respectively, decomposes as
\begin{eqnarray}
  Z_{\alpha\beta}(q)=\sum_in^i_{\alpha\beta}\character{i}\!(q).
\end{eqnarray}
Since we consider the conformal theory living on a torus, introducing boundary 
conditions means breaking up the torus into a cylinder and thus, there are two 
boundaries. With respect to the so called duality condition a physical boundary 
partition function $Z_{\alpha\beta}$ for given boundary conditions $\alpha$ and 
$\beta$ can be written equivalently as:
\begin{eqnarray}
  Z_{\alpha\beta}(q)=\bra{\alpha}
    \tilde q^{\half\left(\Lop_0+\Lbarop_0-\frac{c}{12}\right)}\ket{\beta},
\end{eqnarray}
where $\tilde q\equiv\e^{-2\pi\imag/\tau}$, $\tau$ being the torus parameter. 
Let us introduce another operator $\qop$, remembering $c=-2$:
\begin{eqnarray}
  \qop\equiv q^{\half\left(\Lop_0+\Lbarop_0+\frac{1}{6}\right)}
       =q^{\half\left(\hop+\hbarop+\frac{1}{6}\right)}
       \left[1+\log(q)\cdot\half\Nop+\log(q)^2\cdot\frac{1}{4}\Nop^2\right].
\end{eqnarray}
Here, $q\equiv\e^{2\pi\imag\tau}$.
To verify the last equality one should remember the nilpotency property of 
$\Nop$ and use $\Lop_0+\Lbarop_0=\hop+\hbarop+\Nop$. By equation 
\eqref{eqn:Nsquare_zero} this implies that pairings $\bra{\bs}\qop\ket{\bsc}$ of 
boundary states can contain logarithmic terms proportional of order one at 
maximum, but never of higher order. This is not surprising, since usually, these 
pairings reproduce the torus amplitudes or equivalently, the characters. In 
ordinary conformal field theories the torus amplitudes and the characters span 
exactly the same representation of the modular group. We have two different
representations instead, a three-dimensional one for the physical 
characters and a five-dimensional one presumably for the torus amplitudes 
including the smaller one. Their properties were examined in 
\cite{GaKa:96,GaKa:961,Fl:97,GaKa:98}. 
The main result is that there exists elements of order $\log(q)^1$ in the latter 
representation but neither of them contains $\log(q)^2$ or higher order terms. 
It will turn out that we have to take into account states that are no boundary 
states to reproduce the five-dimensional representation. However, pairings of 
these additional states with the boundary states can contain logarithmic terms 
of order one at maximum as well. Let us calculate the pairings of the boundary 
states first, \ie the cylinder amplitudes. They read as follows:
\begin{eqnarray}
 \begin{array}{ccl@{\qquad}ccl}
  \bra{\VOEbs}\qop\ket{\VOEbs}&\!\!\!=\!\!\!&\character{\V{-1/8}}(q),&
  \bra{\VTEbs}\qop\ket{\VTEbs}&\!\!\!=\!\!\!&\character{\V{3/8}}(q),
  \vspace{2ex} \\
  \bra{\RNbs}\qop\ket{\RNbs}&\!\!\!=\!\!\!&\character{\R{}}(q),&
  \bra{\RObs}\qop\ket{\RObs}&\!\!\!=\!\!\!&\character{\R{}}(q).
 \end{array}
 \label{eqn:bspairing}
\end{eqnarray}  
All other combinations vanish. Particularly, the six states 
$\ket{\VNbs^{\phn}}$, $\ket{\VObs^{\ph{1}}}$, $\ket{\RNObs^{\pm}}$, and 
$\ket{\RONbs^{\pm}}$ are null states with respect to the space spanned by the 
set of boundary states. The characters $\character{i}\!(q)$ were analyzed in 
\cite{GaKa:96,GaKa:961,Ro:96}:
\begin{eqnarray}
  \character{\V{0}}(q)&\!\!\!=\!\!\!&\frac{1}{2\eta(q)}\left(
    \Theta_{1,2}^{\phn}(q)+\left(\del\Theta\right)\!{}_{1,2}^{\phn}(q)\right), 
    \nonumber \\
  \character{\V{1}}(q)&\!\!\!=\!\!\!&\frac{1}{2\eta(q)}\left(
    \Theta_{1,2}^{\phn}(q)-\left(\del\Theta\right)\!{}_{1,2}^{\phn}(q)\right), 
    \nonumber \\
  \character{\V{-1/8}}(q)&\!\!\!=\!\!\!&\frac{1}{\eta(q)}
    \Theta_{0,2}^{\phn}(q), \\
  \character{\V{3/8}}(q)&\!\!\!=\!\!\!&\frac{1}{\eta(q)}
    \Theta_{2,2}^{\phn}(q), \nonumber \\
  \character{\R{}}(q)&\!\!\!\equiv\!\!\!&\character{\R{0}}(q)
    =\character{\R{1}}(q)=\frac{2}{\eta(q)}\Theta_{1,2}^{\phn}(q). 
  \nonumber
\end{eqnarray}
Here, $\eta(q)=q^{1/24}\prod_{n\in N}\left(1-q^n\right)$ is the {\nn Dedekind} 
eta function and $\Theta_{r,2}(q)$ and 
$\left(\del\Theta\right)_{1,2}(q)=\eta(q)^3$ are the ordinary and affine 
{\nn Riemann-Jacobi} theta functions:
\begin{eqnarray}
  \Theta_{r,k}(q)&\!\!\!=\!\!\!&\sum_{n\in Z}q^{\left(2kn+r\right)^2/4k}, 
    \nonumber \\
  \left(\del\Theta\right)_{r,k}\!(q)&\!\!\!=\!\!\!&
    \sum_{n\in Z}\left(2kn+r\right)q^{\left(2kn+r\right)^2/4k}, \\
  \left(\nabla\Theta\right)_{r,k}\!(q)&\!\!\!=\!\!\!&
    \frac{1}{2\pi}\log(q)\left(\del\Theta\right)_{r,k}(q)
    =\imag\tau\left(\del\Theta\right)_{r,k}\!(q). \nonumber
\end{eqnarray}
We also state the characters for $\V{0}$ and $\V{1}$ here as well as the 
logarithmic theta function. They will be needed further on. One learns that the 
boundary states reproduce the three-dimensional representation of the modular 
group, \ie the physical characters. Due to the fact that most of the derived 
boundary states are null states the properties related to the inner structure of 
the indecomposable representations are not visible at this state. Therefore, we 
have to study the boundary states a bit more in detail. In the following we 
especially focus on the structural relations of the states to each other. 
Remember figure \ref{fig:N_eq_delta} where we showed the similarity of $\ndop$ 
in the bulk theory and $\Nop$ for boundary states: Since $\Nop$ has nilpotency 
degree three, the question arises if it is possible to construct a state which 
would not be a boundary state, such that the boundary state $\ket{\RNbs}$ is the 
image of this state under the action of $\Nop$, and the same for $\ket{\RObs}$. 
Unfortunately, it is only possible to find two states $\ket{\Xbs{\index}}$ and 
$\ket{\Ybs{\index}}$ such that
\begin{eqnarray}
  \ket{\Rbs{\index}}=\Nop\ket{\Xbs{\index}}+\ket{\Ybs{\index}}
    \mbox{~~and~~}
  \ket{\Vbs{\index}}=\half\Nop\ket{\Rbs{\index}}=\half\Nop^2\ket{\Xbs{\index}}
    ~~(\index=0,\,1).
\end{eqnarray}
\vspace*{-1.5cm}
\begin{center}
  \begin{picture}(250,120)(-10,-20)
    \put(0,60){\vbox to 0pt{\vss\hbox to 0pt{\hss$\bullet$\hss}\vss}}
    \put(140,60){\vbox to 0pt{\vss\hbox to 0pt{\hss$\bullet$\hss}\vss}}
    \put(230,60){\vbox to 0pt{\vss\hbox to 0pt{\hss$\bullet$\hss}\vss}}
    \put(90,60){\vbox to 0pt{\vss\hbox to 0pt{\hss$\oplus$\hss}\vss}}
    \put(90,0){\vbox to 0pt{\vss\hbox to 0pt{\hss$\bullet$\hss}\vss}}
    \put(5,60){\vector(1,0){75}}
    \put(100,60){\vector(1,0){35}}
    \put(145,60){\vector(1,0){80}}
    \put(90,5){\vector(0,1){50}}
    \put(-7,45){$\ket{\Xbs{\index}}$}
    \put(145,45){$\ket{\Rbs{\index}}$}
    \put(223,45){$\ket{\Vbs{\index}}$}
    \put(83,-15){$\ket{\Ybs{\index}}$}
    \put(40,46){$\Nop$}
    \put(180,46){$\half\Nop$}
  \end{picture}\\[1ex]
\refstepcounter{figure} \label{fig:wbs}
{\em figure \ref{fig:wbs}: weak boundary states}
\ \\[0.5cm]
\end{center}
\noindent
The choice of the states $\ket{\Xbs{\index}}$ and $\ket{\Ybs{\index}}$ is not 
unique. It is possible to add states belonging to the kernel of $\Nop$ to 
$\ket{\Xbs{\index}}$ without changing anything as well as one could subtract 
states from $\ket{\Xbs{\index}}$ that belong to the kernel of $\Nop^2$ and add 
their images under the $\Nop$-operation to $\ket{\Ybs{\index}}$. 
$\ket{\Xbs{\index}}$ and $\ket{\Ybs{\index}}$ generate the boundary states and 
one can therefore call them {\em weak boundary states\/}. This is justified by 
looking at their scalar products with the original boundary states. 
$\ket{\Xbs{\index}}$ and $\ket{\Ybs{\index}}$ can be chosen uniquely in such a 
way that:
\begin{eqnarray}
  \begin{array}{rcl@{\qquad}rcl}
    \bra{\Xbs{\index}}\qop\ket{\Vbs{\index}}&\!\!\!=\!\!\!&
    \character{\V{\index}}(q),&
    \bra{\Xbs{\index}}\qop\ket{\Rbs{\index}}&\!\!\!=\!\!\!&
    \log(q)\cdot\character{\V{\index}}(q), \vspace{1.5ex} \\
    \bra{\Xbs{\index}}\qop\ket{\Ybs{\index}}&\!\!\!=\!\!\!&0,& 
    \bra{\Ybs{\index}}\qop\ket{\Rbs{\index}}&\!\!\!=\!\!\!&
    \character{\R{}}(q)-2\character{\V{\index}}(q), \vspace{1.5ex}\\
    \bra{\Rbs{\index}}\qop\ket{\Rbs{\index}}&\!\!\!=\!\!\!&\character{\R{}}(q).
  \end{array}  
  \label{eqn:torus-ampl}
\end{eqnarray}
We learn that the set of boundary states together with the two states 
$\ket{\Xbs{\index}}$ reproduces the elements of the five-dimensional 
representation of the modular group. Unfortunately, there are terms proportional 
to $\log(q)\Theta_{1,2}(q)$ as well which are not physical and do not belong to
the representation. Luckily, they occur in such a way that they are suppressed 
in certain linear combinations of the boundary states. At this state the 
question remains unanswered in which way the additional states 
$\ket{\Xbs{\index}}$ and $\ket{\Ybs{\index}}$ can physically be interpreted. 
Before moving on to this topic we first want to concentrate on the states 
$\ket{\RNObs^\pm}$ and $\ket{\RONbs^\pm}$ that relate the two generalized 
highest weight representations to each other. Similarly to the definition of the 
states $\ket{\Xbs{\index}}$ there exists states $\ket{\Zbs{01}^\pm}$ and 
$\ket{\Zbs{10}^\pm}$ in such a way that:
\begin{eqnarray}
  \bra{\Zbs{01}^\pm}\qop\ket{\RNObs^\pm}=\bra{\Zbs{10}^\pm}\qop\ket{\RONbs^\pm}
  =\half\character{\R{}}(q).
\end{eqnarray}
These states have the property $\Nop\ket{\Zbs{mn}^\pm}=\ket{\Rbs{mn}^\pm}$, 
$m=1-n$ and fulfill $\Nop^2\ket{\Zbs{mn}^\pm}=0$. They can be interpreted as 
{\em weak boundary states\/} as well.

Boundary states are strongly correlated to propagators that connect the 
holomorphic part (\eg the upper half complex plane, in a very simple setting) to 
the formal anti-holomorphic one (the lower half plane):
\begin{eqnarray}
  \ket{\RNObs^\pm}=\sum_{l,m,n}c^{\,l}_{mn}\ket{l,m}\otimes\overline{\ket{l,n}} 
  &\Leftrightarrow&\Pop_\pm\Udagger\equiv\sum_{l,m,n}c^{\,l}_{mn}
  \ket{l,m}\!\bra{l,n}\mbox{~~and} \nonumber \\
  \ket{\RONbs^\pm}=\sum_{l,m,n}c^{\,l}_{mn}\ket{l,n}\otimes\overline{\ket{l,m}} 
  &\Leftrightarrow&\Pdaggerop_\pm\Udagger\equiv\sum_{l,m,n}c^{\,l}_{mn}
  \ket{l,n}\!\bra{l,m}.
\end{eqnarray}
Here, $\U$ is the anti-unitary operator already introduced above. Because the 
corresponding boundary states satisfy the {\nn Ishibashi} equations 
\eqref{eqn:VirB} and \eqref{eqn:WB} the operators $\Pop_\pm$ and 
$\Pdaggerop_\pm$ commute with the action of the chiral algebra:
\begin{align}
 \begin{split}
  0&=\bra{l_1,a}\otimes\overline{\bra{l_2,b}}
     \left(\Lop_n-\Lbarop_{-n}\right)\ket{\RNObs} \\[1.5ex]
   &=\sum_{l,r,s}\bra{l_1,a}\otimes\overline{\bra{l_2,b}}
     \left(\Lop_n-\Lbarop_{-n}\right)c^{\,l}_{rs}
     \ket{l,r}\otimes\overline{\ket{l,s}} \\
   &=\sum_{l,r,s}c^{\,l}_{rs}\Big\{\bra{l_1,a}\Lop_n\ket{l,r}
     \braket{l,s}{l_2,b}-\braket{l_1,a}{l,r}\bra{l,s}\Lop_n\ket{l_2,b}
     \Big\} \\
   &=\bra{l_1,a}\big[\Lop_n,\Pop\Udagger\big]\ket{l_2,b}  \\[1ex]
   &=\bra{l_1,a}\big[\Lop_n,\Pop\big]\Udagger\ket{l_2,b}.
 \end{split}
\end{align}
We learn that $\Pop$ commutes with the {\nn Virasoro} modes. Analogously one can 
show that $\Pop$ commutes with the modes $\Wop^\spin_n$ and thus, the statement 
is proved. This means that given a boundary state $\ket{\bs}$ the state 
$\Pop\ket{\bs}$ is again a boundary state or equal to zero. It turns out that 
the action of the operators $\Pop$ and $\Pdaggerop$ on the bulk states in the 
representation modules $\R{0}$ and $\R{1}$ is given by:
\begin{eqnarray}
  \begin{array}{rcl@{\qquad}rcl@{\qquad}rcl}
    \Pdaggerop_\pm\ket{\logvac}&\!\!\!=\!\!\!&\ket{\xi^\pm},&
    \Pdaggerop_\pm\ket{\vac}&\!\!\!=\!\!\!&0,& 
    \Pop_+\ket{\logone^\pm}&\!\!\!=\!\!\!&-\ket{\Psi_2^\pm},
    \vspace{1.5ex} \\
    \Pop_\pm\ket{\xi^\mp}&\!\!\!=\!\!\!&\pm\ket{\vac},&
    \Pop_\pm\ket{\one}&\!\!\!=\!\!\!&0,& 
    \Pop_-\ket{\logone^\pm}&\!\!\!=\!\!\!&\ket{\Psi_1^\pm}.
    \vspace{1.5ex} \\    
  \end{array}
\end{eqnarray}
%\begin{eqnarray}
%  \begin{array}{rcl@{\qquad}rcl}
%    \Pdaggerop_\pm\ket{\logvac}&=&\ket{\xi^\pm},&\Pdaggerop_\pm\ket{\vac}&=&0, 
%    \vspace{1.5ex} \\
%    \Pop_\pm\ket{\xi^\mp}&=&\pm\ket{\vac},&\Pop_\pm\ket{\one}&=&0, 
%    \vspace{1.5ex} \\
%    \Pop_+\ket{\logone^\pm}&=&-\ket{\Psi_2^\pm},
%    &\Pop_-\ket{\logone^\pm}&=&\ket{\Psi_1^\pm}.
%  \end{array}
%\end{eqnarray}
Especially, these operators decompose the 
off-diagonal part $\ndop$ of $\Lop_0$:
\begin{eqnarray}
  \ndop=\left\{\begin{array}{ccc}
    \Pop\Pdaggerop&\mbox{on}&\R{0} \vspace{1ex} \\
    \Pdaggerop\Pop&\mbox{on}&\R{1}
  \end{array}\right.\,.
\end{eqnarray}
By use of this equality it is easy to agree on the existence of the 
{\em mixed states\/}:
\begin{eqnarray}
  \begin{array}{ccccc@{\qquad}ccccc}
    \ket{\RNObs^\pm}&\!\!\!=\!\!\!&\Pop_\pm\ket{\RObs}
                    &\!\!\!=\!\!\!&\Pbardaggerop_\pm\ket{\RNbs}, 
   &\ket{\VNbs}&\!\!\!=\!\!\!&\Pbarop_\mp\ket{\RNObs^\pm}
               &\!\!\!=\!\!\!&\Pop_\mp\ket{\RONbs^\pm}, \vspace{1ex} \\
    \ket{\RONbs^\pm}&\!\!\!=\!\!\!&\Pbarop_\pm\ket{\RObs}
                    &\!\!\!=\!\!\!&\Pdaggerop_\pm\ket{\RNbs},
   &\ket{\VObs}&\!\!\!=\!\!\!&\Pdaggerop_\mp\ket{\RNObs^\pm}
               &\!\!\!=\!\!\!&\Pbardaggerop_\mp\ket{\RONbs^\pm}.
  \end{array}
\end{eqnarray}
The action of the operators $\Pop$ and $\Pdaggerop$ and their anti-holomorphic 
partners on the states $\ket{\Xbs{\index}}$ and $\ket{\Ybs{\index}}$ shows that 
they are not independent but that $\ket{\Ybs{\index}}$ can be derived from the 
states $\ket{\Xbs{\index}}$:
\begin{eqnarray}
  \ket{\Ybs{0}}=\Pop\Pbarop\ket{\Xbs{1}} \hspace{0.5cm}\mbox{and}\hspace{0.5cm}
  \ket{\Ybs{1}}=\Pdaggerop\Pbardaggerop\ket{\Xbs{0}}.
\end{eqnarray}
Therefore, the states $\ket{\Xbs{\index}}$ are  the generating states for the
boundary states involving the indecomposable representations. On the other hand, 
we can now justify the denomination {\em weak boundary states\/} in the sense 
that they produce zero under the action of certain operations 
$\hat{{\cal A}}\in\{\Nop,\,\Pop\Pdaggerop\}$:
\begin{eqnarray}
  \hat{{\cal A}}\left(\Xop_n\pm\Xbarop_{-n}\right)\ket{\Xbs{\index}}=0.
\end{eqnarray}
The relations between the boundary states under the action of these operators 
look schematically like the following. On the right hand side we state the 
embedding scheme of the representation $\R{}$ of the local logarithmic conformal 
field theory for the $c=-2$ model 
%analyzed and given in \cite{GaKa:98} 
which looks exactly the same:
%\vspace*{-0.5cm}
%\nopagebreak
\begin{displaymath}
 \begin{array}{l@{\qquad\qquad}r}
  \begin{picture}(140,210)(-92,-50)
    \put(2,40){\vbox to 0pt
        {\vss\hbox to 0pt{\hss$\bullet$\hss}\vss}}
    \put(2,90){\vbox to 0pt
        {\vss\hbox to 0pt{\hss$\bullet$\hss}\vss}}
    \put(62,130){\vbox to 0pt
        {\vss\hbox to 0pt{\hss$\bullet$\hss}\vss}}
    \put(62,0){\vbox to 0pt
        {\vss\hbox to 0pt{\hss$\bullet$\hss}\vss}}
    \put(56,4){\vector(-3,2){48}}
    \put(58,6){\vector(-2,3){52}}
    \put(56,126){\vector(-3,-2){48}}
    \put(58,124){\vector(-2,-3){52}}
    \put(-2,40){\vbox to 0pt
        {\vss\hbox to 0pt{\hss$\bullet$\hss}\vss}}
    \put(-2,90){\vbox to 0pt
        {\vss\hbox to 0pt{\hss$\bullet$\hss}\vss}}
    \put(-62,130){\vbox to 0pt
        {\vss\hbox to 0pt{\hss$\bullet$\hss}\vss}}
    \put(-62,0){\vbox to 0pt
        {\vss\hbox to 0pt{\hss$\bullet$\hss}\vss}}
    \put(-8,36){\vector(-3,-2){48}}
    \put(-6,84){\vector(-2,-3){52}}
    \put(-8,94){\vector(-3,2){48}}
    \put(-6,46){\vector(-2,3){52}}
    \put(55,0){\vector(-1,0){110}}
    \put(55,130){\vector(-1,0){110}}
    \put(-67,-15){\hbox to 0pt{\hss$\ket{\VNbs}$}}
    \put(67,-15){\hbox to 0pt{$\ket{\RNbs}$\hss}}
    \put(-67,135){\hbox to 0pt{\hss$\ket{\VObs}$}}
    \put(67,135){\hbox to 0pt{$\ket{\RObs}$\hss}}
    \put(-12,18){\hbox to 0pt{$\ket{\RNObs^\pm}$\hss}}
    \put(-12,107){\hbox to 0pt{$\ket{\RONbs^\pm}$\hss}}
    \put(0,-35){\hbox to 0pt{\hss{\em boundary states}\hss}}
  \end{picture}
 &
  \begin{picture}(170,210)(-92,-50)
    \put(2,40){\vbox to 0pt
        {\vss\hbox to 0pt{\hss$\bullet$\hss}\vss}}
    \put(2,90){\vbox to 0pt
        {\vss\hbox to 0pt{\hss$\bullet$\hss}\vss}}
    \put(62,130){\vbox to 0pt
        {\vss\hbox to 0pt{\hss$\bullet$\hss}\vss}}
    \put(62,0){\vbox to 0pt
        {\vss\hbox to 0pt{\hss$\bullet$\hss}\vss}}
    \put(56,4){\vector(-3,2){48}}
    \put(58,6){\vector(-2,3){52}}
    \put(56,126){\vector(-3,-2){48}}
    \put(58,124){\vector(-2,-3){52}}
    \put(-2,40){\vbox to 0pt
        {\vss\hbox to 0pt{\hss$\bullet$\hss}\vss}}
    \put(-2,90){\vbox to 0pt
        {\vss\hbox to 0pt{\hss$\bullet$\hss}\vss}}
    \put(-62,130){\vbox to 0pt
        {\vss\hbox to 0pt{\hss$\bullet$\hss}\vss}}
    \put(-62,0){\vbox to 0pt
        {\vss\hbox to 0pt{\hss$\bullet$\hss}\vss}}
    \put(-8,36){\vector(-3,-2){48}}
    \put(-6,84){\vector(-2,-3){52}}
    \put(-8,94){\vector(-3,2){48}}
    \put(-6,46){\vector(-2,3){52}}
    \put(55,0){\vector(-1,0){110}}
    \put(55,130){\vector(-1,0){110}}
    \put(-67,-15){\hbox to 0pt{\hss$\vac$}}
    \put(67,-15){\hbox to 0pt{$\logvac$\hss}}
    \put(-67,135){\hbox to 0pt{\hss$\one$}}
    \put(67,135){\hbox to 0pt{$\logone$\hss}}
    \put(-7,25){\hbox to 0pt{$\rho^{\,\pm}$\hss}}
    \put(-7,100){\hbox to 0pt{$\bar\rho^{\,\pm}$\hss}}
    \put(0,-35){\hbox to 0pt{\hss$\R{}$\hss}}
  \end{picture}
 \end{array}
\end{displaymath}
\refstepcounter{figure} \label{fig:R01}
\vspace*{-4ex}
\begin{center} {\em figure \ref{fig:R01}: boundary states vs.\ local theory}
\end{center}
The perfect one-to-one correspondence between these two diagrams suggests that 
there is a deeper relation between the local theory and the one with boundaries.
This is very interesting, especially because the boundary states were derived
completely independent to the local theory. One the other hand, that there
has to be at least a link between the two theories is already clear from
the very beginning. The local theory fuses together a chiral and an 
anti-chiral copy of the rational $c=-2$ theory. To keep locality, certain
states have to be devided out, namely the image of $\Lop_0-\Lbarop_0$. 
Equation \eqref{eqn:VirB} treated for $n=0$ states that exactly 
the states with non-vanishing norm in the range of $\Lop_0-\Lbarop_0$ are not 
allowed to contribute to the boundary states.

%-----------------------------------------------------------
% section 5: Partition function
%-----------------------------------------------------------
\section{Partition function}

Looking at (\ref{eqn:bspairing}) and taking only the non-null states into 
account, one immediately finds the three-dimensional representation of the 
modular group in this case. Let us try to apply {\nn Cardy}'s method to obtain 
the physically relevant boundary conditions. For the set 
$\{\character{\R{0}},\,\character{\R{1}},\,\character{\V{-1/8}},\,
\character{\V{3/8}}\}$, the ${\cal S}$ and ${\cal T}$ matrices that give the 
transformations of the characters under the modular transformations 
$\tau\rightarrow 1/\tau$ and $\tau\rightarrow \tau+1$ respectively are dealt 
with in \cite{Ro:96}. There are six proper choices due to the fact that there 
are four independent representation modules whose characters form only a 
three-dimensional representation of the modular group. One of the possibilities 
is:
\begin{eqnarray}
  \mathcal{S}=\left(
    \begin{array}{cccc}
      \phm\frac{\imag}{2}&-\frac{\imag}{2}&\phm\frac{1}{4}&-\frac{1}{4}
      \vspace{0.1cm}\\
      -\frac{\imag}{2}&\phm\frac{\imag}{2}&\phm\frac{1}{4}&-\frac{1}{4}
      \vspace{0.1cm}\\
      \phm 1&\phm 1&\phm\half&\phm\half  \vspace{0.1cm}\\
      -1&-1&\phm\half&\phm\half
    \end{array}\right),\hspace{0.2cm}
  \mathcal{T}=\left(
    \begin{array}{cccc}
      0&\e^{\imag\pi/6}&0&0  \vspace{0.1cm}\\
      \e^{\imag\pi/6}&0&0&0  \vspace{0.1cm}\\
      0&0&\e^{-\imag\pi/12}&0  \vspace{0.1cm}\\
      0&0&0&-\e^{-\imag\pi/12}
    \end{array}\right).    
\end{eqnarray}
The associated charge conjugation matrix ${\cal C}$ is the permutation matrix 
that permutes the first two lines. It reads:
\begin{eqnarray}
  \mathcal{C}=\left(
    \begin{array}{cccc}
      0&1&0&0 \vspace{0.1cm}\\
      1&0&0&0 \vspace{0.1cm}\\
      0&0&1&0 \vspace{0.1cm}\\
      0&0&0&1
    \end{array}
  \right).
\end{eqnarray}      
Altogether, these matrices satisfy
$\mathcal{S}^4=1$ and $(\mathcal{S}\mathcal{T})^3=\mathcal{S}^2=\mathcal{C}$.
Taking the set of boundary basis states to be 
$\{\ket{\RNbs},\,\ket{\RObs},\,\ket{\VOEbs},\,\ket{\VTEbs}\}$ we arrive at the 
starting point for {\nn Cardy}'s method. Firstly, we have to construct the 
vacuum boundary state $\ket{\bf \logvac}$ which can be written in terms of the 
basis states as
\begin{eqnarray}
  \ket{\bb{\logvac}}=a\ket{\RNbs^{\phn}}+b\ket{\RObs^{\phn}}
                    +c\ket{\VOEbs}+d\ket{\VTEbs}.
\end{eqnarray}
The remaining problem is that $\mathcal{S}^j_0$ is required to be positive 
valued. This is obviously not the case, the coefficients are even complex. We 
therefore introduce the conjugate vacuum representation boundary state 
$\ket{\bf \logvac^\vee}\,$:
\begin{eqnarray}
  \ket{\bb{\logvac}^\vee}=a^*\ket{\RNbs^{\phn}}+b^*\ket{\RObs^{\phn}}
                         +c^*\ket{\VOEbs}+d^*\ket{\VTEbs},
\end{eqnarray}
and look at boundary conditions of the form $(\alpha^\vee,\beta)$, \ie on the 
one hand side of the cylinder we apply the condition $\alpha^\vee$ instead of 
$\alpha$ and on the other hand the condition $\beta$. With this, it is an 
easy task to calculate the boundary states \ala\/ {\nn Cardy}:
\begin{eqnarray}
  \ket{\bb{i}}=\sum_j\frac{S_i^j}{\sqrt{S_0^j}}\ket{j}.
\end{eqnarray}
Here, $\ket{j}$ denotes the boundary basis state belonging to the representation 
$j$ and $\ket{\bb{i}}$ labels the physical relevant boundary state that 
corresponds to a bulk {\nn Hamiltonian} that contains only the representation 
$i$ in its spectrum. The physical boundary conditions finally read:
\begin{eqnarray}
  \begin{array}{rcl}
  \ket{\bb{\logvac}}&\!\!\!=\!\!\!&
     \frac{1}{\sqrt{2}}\e^{\,\imag\pi/4}\ket{\RNbs}
    -\frac{1}{\sqrt{2}}\e^{\,-\imag\pi/4}\ket{\RObs}
    +\half\ket{\VOEbs}+\frac{i}{2}\ket{\VTEbs}, \vspace{2ex}\\
  \ket{\bb{\logone}}&\!\!\!=\!\!\!&
    -\frac{1}{\sqrt{2}}\e^{\,\imag\pi/4}\ket{\RNbs}
    +\frac{1}{\sqrt{2}}\e^{\,-\imag\pi/4}\ket{\RObs}
    +\half\ket{\VOEbs}+\frac{\imag}{2}\ket{\VTEbs}, \vspace{2ex}\\
  \ket{\bb{\mu}}    &\!\!\!=\!\!\!&
     \sqrt{2}\e^{\,-\imag\pi/4}\ket{\RNbs}-\sqrt{2}\e^{\,\imag\pi/4}\ket{\RObs}
    +\ket{\VOEbs}-\imag\ket{\VTEbs}, \vspace{2ex}\\
  \ket{\bb{\nu}}    &\!\!\!=\!\!\!&
    -\sqrt{2}\e^{\,-\imag\pi/4}\ket{\RNbs}+\sqrt{2}\e^{\,\imag\pi/4}\ket{\RObs}
    +\ket{\VOEbs}-\imag\ket{\VTEbs}. 
  \end{array}
  \label{eqn:3dbs}
\end{eqnarray}
Here, the states are labeled in correspondence to the cyclic states of the 
underlying bulk representation. The conjugate states are given by complex 
conjugation of the coefficients. The boundary states \eqref{eqn:3dbs} are not 
uniquely defined but rather chosen up to a $\mathbb{Z}_4$ symmetry in the 
coefficient phases. Of course, it is as well possible to start from any of the 
other five proper definitions of the ${\cal S}$ and ${\cal T}$ matrices which 
lead to the same solutions. By construction it is clear that the partition 
function coefficients with respect to these states are equal to the fusion 
rules that are related to the elements of the ${\cal S}$ matrix by the 
{\nn Verlinde} formula \cite{Ve:88}:
\begin{eqnarray}
  n_{i^\vee j}^k=N_{ij}^k=
  \sum_r\frac{{\cal S}_r^i{\cal S}_r^j{\cal S}_k^r}{{\cal S}_0^r}.
  \label{eqn:Verlinde}
\end{eqnarray}
Usually, the ${\cal S}$ matrix diagonalizes the fusion rules. As indicated in 
\cite{Ro:96} this is not the case. Instead, the fusion matrices are transformed 
into block-diagonal form.

To summarize the result of this section we were able to show that the standard
{\nn Cardy} procedure works perfectly well in the $c=-2$ theory on the character 
representation of the modular group. It seems natural that this could be generalized
to other such theories.

%-----------------------------------------------------------
% section 6: The five-dimensional representation
%-----------------------------------------------------------
\section{The five-dimensional representation}

Now, we want to focus on the five-dimensional representation of the modular 
group. To do so, we study the complete set of boundary states plus the two 
generating states $\ket{\Xbs{\index}}$. The representation was investigated 
in \cite{GaKa:96,GaKa:961,GaKa:98,Fl:97}. In 
\cite{Fl:97}, an approach based on ideas of {\nn S.\,D.~Mathur} \etal 
\cite{MaMuSe:88,MaMuSe:89} was taken, which we will follow here: 
Let the linearly independent set of characters be given by the set
\begin{eqnarray}
  \Big\{\character{\V{0}},\,\character{\V{-1/8}},\,\character{\V{1}},
  \character{\V{3/8}},\,2\character{\widetilde{\R{}}}\equiv
    \frac{2}{\eta}\left[\Theta_{1,2}^{\phn}
      +\imag\alpha\left(\nabla\Theta\right)\!{}_{1,2}^{\phn}\right]\Big\},
 \end{eqnarray}
where $\alpha\in \mathbb{R}$ is arbitrary. The corresponding ${\cal S}$ matrix 
that transforms the characters under $\tau\rightarrow -1/\tau$ reads:
\begin{eqnarray}
  \mathcal{S}=\left(\begin{array}{ccccc}
    \phm\frac{1}{2\alpha}&\frac{1}{4}&\phm\frac{1}{2\alpha}&-\frac{1}{4}&
    -\frac{1}{4\alpha} \vspace{0.2cm}\\
    \phm 1&\half&\phm 1&\phm\half&\phm 0 \vspace{0.2cm}\\
    -\frac{1}{2\alpha}&\frac{1}{4}&-\frac{1}{2\alpha}&-\frac{1}{4}&
    \phm\frac{1}{4\alpha} \vspace{0.2cm}\\
    -1&\half&-1&\phm\half&\phm0 \vspace{0.2cm}\\
    -2\alpha&1&\phm 2\alpha&-1&\phm0
    \end{array}\right).
\end{eqnarray}
In order to find the elements of the five-dimensional representation we had to
introduce partners to the boundary null-states that themselves are no boundary 
states and serve as duals for the null-states. However, we want to have the 
possibility that at least in a physical limit these states vanish. To do so 
let us not consider the states $\ket{\Vbs{\index}}$ and their corresponding 
bra-states $\bra{\Xbs{\index}}$ but instead use the two renormalized states
\begin{eqnarray}
  \ket{\Vbs{\index}}\longrightarrow\frac{2\pi}{\sqrt{2}\alpha}\ket{\Vbs{\index}}
  \mbox{~~and~~}
  \bra{\Xbs{\index}}\longrightarrow \frac{\sqrt{2}\alpha}{2\pi}\bra{\Xbs{\index}}.
\end{eqnarray}
It is obvious that the pairings $\bra{\Xbs{\index}}\qop\ket{\Vbs{\index}}$ do not change 
for any choice of $\alpha$, in particular for $\alpha=2\pi/\sqrt{2}$ for which 
we obtain the original states and that, on the other hand, the pairings 
$\bra{\Xbs{\index}}\qop\ket{\Rbs{\index}}$ get an additional pre-factor
$(\sqrt{2}\alpha)/(2\pi)$ such that we obtain
\begin{eqnarray}
  \bra{\Xbs{\index}}\qop\ket{\Rbs{\index}}=
    \frac{\sqrt{2}\alpha}{2\pi}\log(q)\character{\V{\index}}(q)
    \mbox{~~and~~}\bra{\Xbs{\index}}\qop\ket{\Vbs{\index}}=\character{\V{\index}}(q).
\end{eqnarray}
Following the {\nn Cardy} formalism the physical vacuum boundary state is up to 
a choice of phases in ordinary conformal field theories given by
\begin{eqnarray}
  \ket{\bb{\vac}}=\sum_j\sqrt{{\cal S}_0^j}\,\ket{j}.
\end{eqnarray}
Here, $j$ runs as in the previous section over the representation modules and 
the corresponding boundary basis states. In our case, this has to be treated 
more carefully. First of all, since the concerned elements of the ${\cal S}$ 
matrix are not positive we again have to introduce a conjugate vacuum 
representation $\ket{\bb{\vac}^\vee}$ in order to be able to follow 
{\nn Cardy}'s argumentation. If we naively compute the boundary states we 
find that they do not exactly reproduce the characters for which the ${\cal S}$ 
matrix is written down in the above form. Recall that we have the states 
$\ket{\Zbs{\index(1-\index)}}$ at our disposal. By putting together all these 
arguments and 
following the standard formalism one can write down the following boundary 
states:
\begin{eqnarray}
  \ket{\bb{\vac}}&\!\!\!=\!\!\!&\frac{1}{2\sqrt{\alpha}}\Big\{
     \sqrt{2}\left(\ket{\VNbs}+\ket{\VObs}\right)
     +\imag\left(\ket{\RNbs}-\ket{\RObs}\right) \nonumber \\[-0.1cm]
  && +\left(\ket{\RNObs}+\ket{\RONbs}\right)\Big\} 
     +\half\left(\ket{\VOEbs}+\imag\,\ket{\VTEbs}\right),\nonumber \\
  \ket{\bb{\mu}}&\!\!\!=\!\!\!&\sqrt{2\alpha}\left\{\ket{\VNbs}
     +\ket{\VObs}\right\}
     +\left(\ket{\VOEbs}-\imag\,\ket{\VTEbs}\right), \nonumber \\
  \ket{\bb{\one}}&\!\!\!=\!\!\!&\frac{1}{2\sqrt{\alpha}}\Big\{
     -\sqrt{2}\left(\ket{\VNbs}+\ket{\VObs}\right)
     +\imag\left(-\ket{\RNbs}+\ket{\RObs}\right) \nonumber \\[-0.1cm]
  && +\left(-\ket{\RNObs}+\ket{\RONbs}\right)\Big\}
     +\half\left(\ket{\VOEbs}+\imag\,\ket{\VTEbs}\right),  \\
  \ket{\bb{\nu}}&\!\!\!=\!\!\!&-\sqrt{2\alpha}\left\{\ket{\VNbs}
     +\ket{\VObs}\right\}
     +\left(\ket{\VOEbs}-\imag\,\ket{\VTEbs}\right), \nonumber \\
  \ket{\bb{\logvac}}&\!\!\!=\!\!\!&2 \sqrt{2\alpha^3}\Big\{
     -\ket{\VNbs}+\ket{\VObs}\Big\}
     +2\left(\ket{\VOEbs}+\imag\,\ket{\VTEbs}\right), \nonumber
%  \ket{\bb{\logone}}&\!\!\!=\!\!\!&\sqrt{\alpha x}\Big\{
%     \sqrt{2}\alpha x\left(-\ket{\VNbs}+\ket{\VObs}\right)
%     +\imag\left(\ket{\RNbs}+\ket{\RObs}\right) \nonumber \\[-0.1cm]
%  && -a\ket{\RNObs}-b\ket{\RONbs}\Big\} 
%     +\left(\ket{\VOEbs}+\imag\,\ket{\VTEbs}\right). \nonumber		     
\end{eqnarray}  
First of all, the coefficients for the six boundary basis states that belong to 
only one representation module each fulfill the {\nn Cardy} law, \ie are up to
the phases given by ${\cal S}_k^j/\sqrt{{\cal S}_0^j}$. Note that the two states
corresponding to the indecomposable representations are treated by the same matrix
elements up to a phase change. Secondly, one recognizes 
terms proportional to the mixed states that are added by hand. Their presence is
explained as they happen to be counter terms in order to get the partition 
function coefficients to satisfy the {\nn Verlinde} formula as in the ordinary 
cases. They appear due to the fact that the characters are not what one gets out 
of the pairings. Nevertheless, the representation of the modular group is the 
same, of course. Indeed, by calculating the partition functions for given 
boundary conditions $(i^\vee,j)$ it turns out that
\begin{eqnarray}
  Z_{i^\vee j}^{\phn}(q)=\sum_k n_{i^\vee j}^k\chi_k^{\phn}(q)
\end{eqnarray}
with $n_{i^\vee j}^k$ being {\em nearly\/} equal to the fusion coefficients 
derived from the given ${\cal S}$ matrix via the {\nn Verlinde} formula 
\eqref{eqn:Verlinde}. These fusion coefficients are equal to the physical ones 
up to some identifications, \ie $2\V{0}+2\V{1}\equiv\R{0}\equiv\R{1}$ 
concerning the number of states on each level and in the physical limit 
$\alpha\longrightarrow 0$ under which the contributions of the non-boundary 
states and the mixed states vanish. The problem with the partition function is 
that one finds terms proportional to $\log(q)\Theta_{1,2}(q)$. Fortunately, they 
come with a pre-factor $\alpha$ and thus vanish in the limit 
$\alpha\longrightarrow 0$. One even finds that under this limit 
$n_{i^\vee j}^k=N_{ij}^k$. Again, the fusion matrices are not diagonalized by 
the ${\cal S}$ matrix but instead transformed in block-diagonal form. However,
it is not possible to apply this limit to the boundary states themselves since 
they get divergent. This is related to the observation in \cite{Fl:97} of the 
same problem occurring for the ${\cal S}$ matrix, namely that even though it 
reproduces the fusion coefficients in the limit $\alpha\longrightarrow 0$, it is 
not possible to apply this limit to the matrix itself due to the fact that in 
this limit the set of characters gets linearly dependent. 

%-----------------------------------------------------------
% section 7: Discussion
%-----------------------------------------------------------
\section{Discussion}

We presented a mathematically consistent way on how to treat boundary states in
logarithmic conformal field theories and applied it to the rational $c=-2$ 
logarithmic conformal field theory. The advantage of the invented method is it's 
simplicity: We only make use of an arbitrary basis for each representation 
module and the expansion of a given state with respect to this basis. In 
particular, this basis does not have to be orthonormal. As a side-effect the 
algorithm derives the inverse metric on each representation module. The algorithm
turns out to be finite in the sense that the components of the boundary states can 
be derived up to any given finite level in a finite number of steps.

We could identify ten states that obey the {\nn Ishibashi} boundary state 
conditions and that can be arranged in a scheme very similar to the embedding 
scheme of the local theory proposed by {\nn M.\,R.~Gaberdiel} and 
{\nn H.\,G.~Kausch}. Six of these states turned out to be null states in the 
space of boundary states. The remaining four together with the corresponding
${\cal S}$ matrix can be treated by the standard {\nn Cardy} formalism in order
to obtain the physical relevant boundary conditions concerning the 
three-dimensional representation of the modular group. On the other hand, 
we could identify additional states in such a way that their pairings with the 
boundary states together with the non-vanishing boundary state-boundary state
pairings reproduce the five-dimensional representation of the modular group. By 
referring to these additional states as the dual states corresponding to the 
boundary null-states we were able to apply a slightly modified version of the 
{\nn Cardy} formalism in this case and obtained at least in a physical limit the 
wanted relation between the partition function coefficients and the fusion rules 
of the bulk theory. For the application of this limit, the same problems arise 
as for the ${\cal S}$ matrix that transforms the bulk characters. It is 
remarkable that the {\nn Cardy} formalism works in both cases. The meaning of 
this, however, is still unknown but it is worth noting in this context that we
could identify exactly the same elements that are presumed to form the set
of torus amplitudes in the bulk theory. The investigation of the deeper meaning
of the additional so-called weak boundary states is left for future work.

In ordinary rational conformal field theories, solving {\nn Cardy}'s
consistency condition reduces to finding non-negative integer-valued matrix
representations (so-called NIM-representations) of the {\nn Verlinde} algebra 
\cite{BePePeZu:00}, see also \cite{FuRuSch:01}. Notice that 
our five-dimensional solution for the $c=-2$ logarithmic 
conformal field theory does involve negative integers. These occur in
a very similar fashion as in the computation of fusion matrices along
the lines of \cite{Fl:97}. This demonstrates that logarithmic conformal
field theories, although they share many properties with rational conformal
field theories, cannot entirely be put on equal footing with them.
However, the negative integers are not as bad as they initially may
appear, since they precisely reflect the linear dependencies among the
boundary states which appear in the above discussed limiting procedure. If
these dependencies are taken into account in the correct way, the final
solution can be written without negative integers. Unfortunately, this
last step has to be done by hand, since the computation via the
${\cal S}$ matrix and {\nn Cardy}'s ansatz inevitabely will lead to some 
negative integer coefficients. It remains an interesting open question, 
in which sense more general solutions than the NIM-representations should 
be taken into account for settings slightly more general than ordinary rational
conformal field theory.

{\nn Y.~Ishimoto} conjectured that for every indecomposable representation of 
rank 2, there exists exactly one boundary state \cite{Ism:01}. Our analysis 
shows that this conjecture holds in the $c=-2$ case, even though not strictly. 
We derived two boundary states for the indecomposable representations 
each where one only refers to the contained subrepresentation, respectively. 
This seems in contradiction to the stated conjecture. On the other hand, one 
of these two states turns out to be a null state in the space of boundary 
states.

In one of the first works on this topic, {\nn I.\,I.~Kogan} and 
{\nn J.\,F.~Wheater} tried to fix the zero-norm state problem by a perturbative 
procedure. By doing this, they introduced a physical limit as well that looks 
much like our's, namely they multiplied the vacuum {\nn Ishibashi} boundary 
state by a factor of $1/\epsilon$ and did the limiting in the calculation of the 
pairings. The trouble they ran into is that the characters they arrive at are 
not the ones that are really observed. A more severe problem is based in the 
perturbation process, namely if one introduces a non-vanishing scalar product of 
the bulk vacuum state with itself, then the $\Lop_0$ mode does not behave well 
any longer, \ie the {\nn Shapovalov} forms would turn out to be non-symmetric. 
Nevertheless, the principle idea of introducing such a limit is still the 
same.

{\nn S.~Kawai} and {\nn J.\,F.~Wheater} tried to solve the boundary problem by 
introducing symplectic fermions. With this, they found six boundary states and 
were able to relate them to the bulk properties in the usual way by defining the 
bra and ket states completely independent of each other. Another issue was 
that they could either choose a set of states that corresponded to the local 
theory or to the chiral theory.

None of the cited works, however, discussed the mixed states that intertwine the 
two 
different indecomposable bulk representations. The existence of these states is 
justified by the comparison to the local theory. There is still some work to do, 
but we conjecture that there has to be a deeper fundamental relation between the 
boundary and the local theory.

\bigskip
\noindent{\bf Acknowledgement:} The work of M.F.\ is supported by the
DFG string network (SPP no.\ 1096), Fl 259/2-1.

%\vspace*{-2pt}

\end{document}